\documentclass[aps,
footinbib,
superscriptaddress,
showpacs,twocolumn,
pra]{revtex4-1}

\usepackage{bm}
\usepackage{graphicx}
\usepackage{latexsym}
\usepackage{amssymb}
\usepackage{amsmath}
\usepackage{color}

\begin{document}
\bibliographystyle{apsrev}

\title{Charge dynamics of a molecular ion immersed in a Rydberg-dressed atomic lattice gas}

\author{Rick Mukherjee}
\affiliation{Department of Physics, Indian Institute of Science Education and Research, Bhopal, India}
\affiliation{Department of Physics, Imperial College, SW7 2AZ, London, UK}

\date{\today}
\begin{abstract}
Charge dynamics in an ultra-cold setup involving a laser dressed atom and an ion is studied here. This transfer of charge is enabled through molecular Rydberg states that are accessed via a laser. The character of the charge exchange crucially depends on the coupling between the electronic dynamics and the motional dynamics of the atoms and ion. The molecular Rydberg states are characterized and a criterion for distinguishing coherent and incoherent regimes is formulated. Furthermore the concept is generalized to the many-body setup as the ion effectively propagates through a chain of atoms. Aspects of the transport such as its direction can be controlled by the excitation laser. This leads to new directions in the investigation of hybrid atom-ion systems that can be experimentally explored using optically trapped strontium atoms.
\end{abstract}

\maketitle

Ultracold atoms in optical lattices opened the door for experimental studies of a wide range of quantum many-body problems \cite{Jaksch, Greiner, Lewenstein, Bloch}. Similar breakthroughs have been achieved in trapped ion systems \cite{Cirac, Blatt}. One of the main motivations of using such systems is to simulate spin models \cite{Britton,Islam} in a controlled environment including Hamiltonians which are intractable by conventional numerical methods \cite{Roy}. Emerging from these efforts, there is a growing interest in exploring hybrid systems formed of trapped atoms and ions \cite{schmid, Rellergert, Goold10, Denschlag14,Secker16,Krukow16,Tomza,Haze1}. This combination enables access to a plethora of novel phenomena such as strongly coupled polaron states \cite{Kalas,Cucchietti,Casteels}, long-range collisions \cite{grier,Ratschbacher,Idziaszek}, study of exciton transport \cite{Wuester}, electron-phonon coupling in Fermi gases \cite{Bissbort}, many-body quantum dynamics \cite{Schurer14,Schurer15,Schurer16}, implementation of atom-ion quantum gates \cite{Doerk}, switches for information transfer \cite{Gerritsma},  quantum simulation of novel ultra-cold chemistry \cite{Deiglmayr} as well as the formation of mesoscopic molecular ions \cite {cote1,Schurer17}.

Charge exchange are processes central in atom-ion systems and has relevance in the study of chemical reactions \cite{Ratschbacher,Zygelman,Saito,Sikorsky} as well as charge transport in the ultra-cold domain \cite{cote2}. Resonant charge exchange in atom-ion setups plays a crucial role in the cooling of ions \cite{Ravi,Dutta,Haze2,Meir}. At sufficiently low temperatures, the mechanism for charge exchange involves electron hopping from neutral atoms to a neighbouring ion. This process is highly suppressed for ground state atoms due to the negligible overlap between the electron wave function with the orbital of a nearby ion.  This unfavourable situation can change for highly excited (Rydberg-)atoms \cite{Gallagher} where the large spatial extent of the electronic wave function enhances the probability for electron hopping onto the ion \cite{Lesanovsky}.
\begin{figure}[t!]
	\includegraphics[width=1.0\columnwidth]{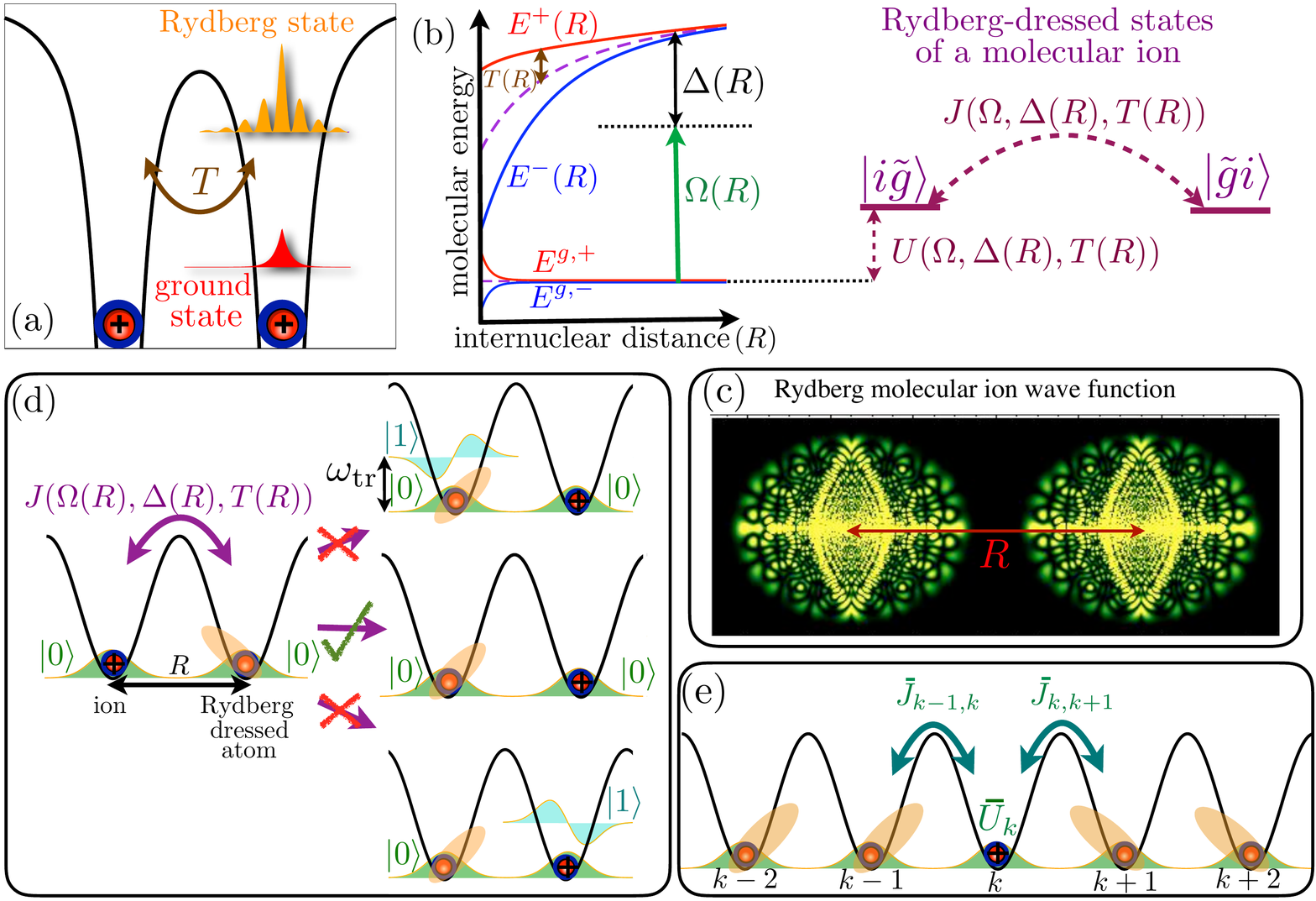}
	\caption{(a) Illustration of the key principle: While a low-lying state (shown in red) remains localized, a Rydberg state (shown in orange) can tunnel through the ionic potential barrier (black lines) at rate $T$. (b) A laser addresses the excited Rydberg molecular states with effective coupling $\Omega(R)$ and detuning $\Delta(R)$. Tunneling ($T(R)$) is given by the splitting between the gerade ($E^{+}(R)$) and ungerade ($E^{-}(R)$) states which are in fact the electronic molecular ion states. In the Rydberg dressed picture, $J(R)$ is the effective hopping and $U(R)$ is the overall light shift of the relevant Rydberg-dressed states (denoted by $|i\tilde{g}\rangle$ and $|\tilde{g}i\rangle$). (c) Depicts the probability density for a particular  Rydberg-dressed molecular ion wave function. (d) Coherent dynamics is facilitated by confining the ion and the Rydberg-dressed atom in an identical double well optical trap. Initially prepared in their motional ground states $|0\rangle$ (shown in green), the coupling to higher motional states such as $|1\rangle$ (shown in aquamarine) are suppressed by choosing optimum optical and trapping conditions. (e) The two particle picture is generalized to obtain the effective many-body charge transport model with nearest neighbour hopping $\bar{J}_{k,k\pm1}$ and on-site energy $\bar{U}_k$.} \label{fig1}
\end{figure}

The aim of this work is to investigate the charge dynamics in an atom-ion hybrid system that is formed by a deep optical lattice filled with a single atom per site out of which one is ionized. Before moving to the many-body problem, we first study the two body problem involving the atom and an ion. The underlying key ingredient is the existence of electronic molecular Rydberg states that encompass the ion and an adjacent atom. For a Rydberg atom, the wave function of the electron has a large spatial extension which enhances the hopping probability as depicted in Fig.~\ref{fig1}(a). The tunneling rate is determined by the splitting between the corresponding potential curves of opposite symmetry as shown in Fig.~\ref{fig1}(b). However, enhanced tunneling occurs at inter-nuclear distances where the Rydberg atom is considerably polarised by the ion leading to strong l-mixing of the Rydberg states resulting in the formation of complex Rydberg molecular ion states as shown in Fig.~\ref{fig1}(c). In the presence of a detuned laser, the coupling of the ground state atom to Rydberg states is described by a dressed atom picture and the ion dynamics is effectively given by $J(R)$. The implicit dependence of $J(R)$ on inter-atomic distance entangles the electronic and motional dynamics which then introduces de-coherence into the dynamics. Coherent charge dynamics can be achieved by trapping the atoms and the ion in an identical potential [see  Fig.~\ref{fig1}(d)] similar to \cite{Mukherjee}. The natural extension of the two particle picture to the many-body system leads to de-localised charge dynamics which is interesting in its own right. However, in certain regimes of the parameter space, the many-body charge dynamics is effectively dictated with nearest neighbour hopping as shown schematically in Fig.\ref{fig1}(e).
\begin{figure}
	\includegraphics[width=1.0\columnwidth]{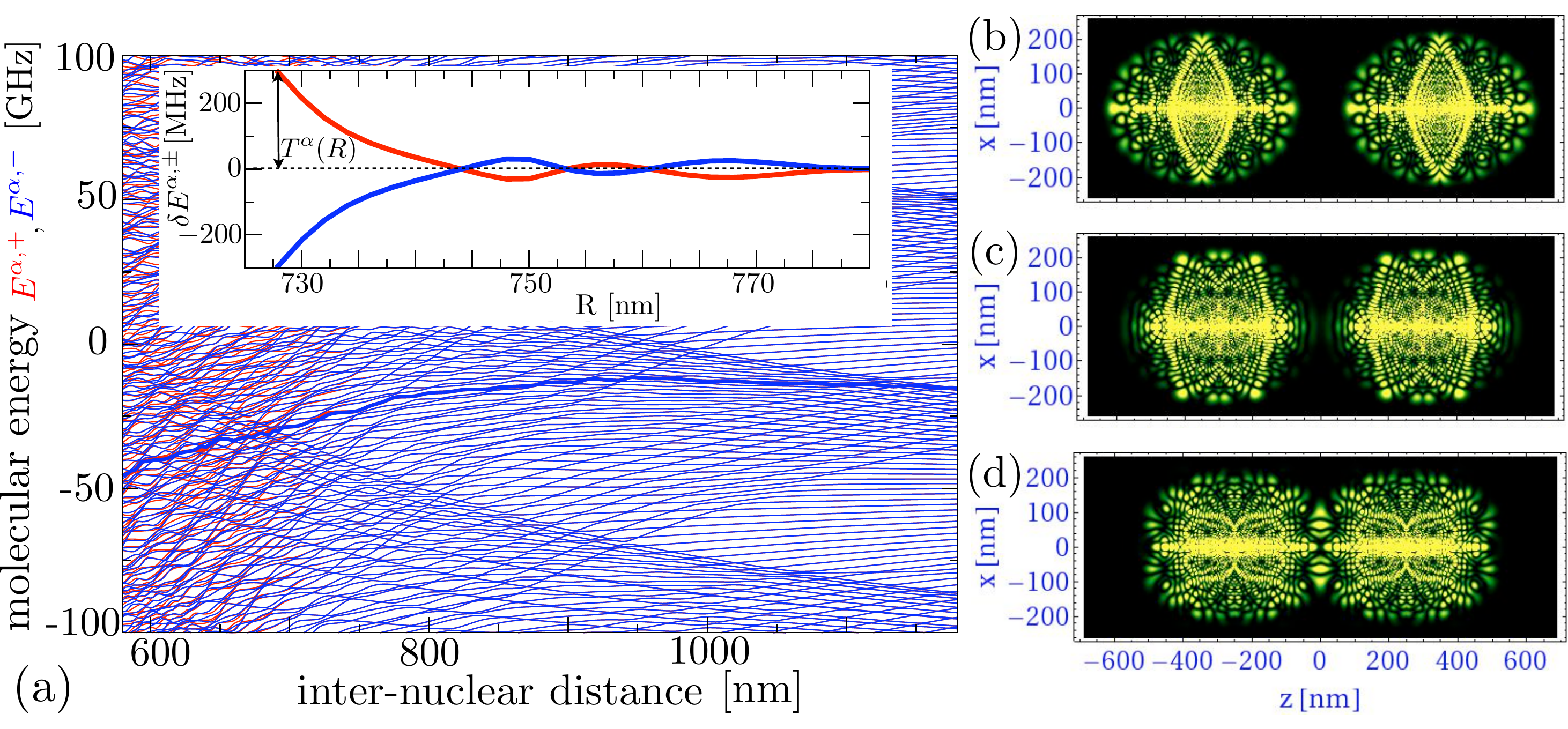}
	\caption{(a) Potential curves for high lying Rydberg states of a ${\rm Sr}_2^{+}$ molecular ion. The molecular energies are given relative to the ${\rm Sr}_2^{+}(50S)$ asymptote. The relative energies, $\delta E^{\alpha,(\pm)}=E^{\alpha,(\pm)}-(E^{\alpha,+}+E^{\alpha,-})/2$, of a selected pair ($\alpha=243$) with (un)gerade symmetry is shown in the inset for which the tunnel splitting is as large as several $100~{\rm MHz}$ at an internuclear distance of $R=730~{\rm nm}$. (b-d) Plotting the complex Rydberg molecular ion wave functions for selected pair of potentials shown in the inset for three different inter-nuclear distances, 700 nm in (b), 670 nm in (c) and 640 nm in (d).}
	\label{fig2}
\end{figure}

\section{Rydberg Molecular Ion} 
The Rydberg molecular ion states are calculated for strontium (Sr) by adopting a linear combination of localized orbitals. The orbitals ($\psi_{nl}$) correspond to Rydberg states of Sr obtained using a single active electron approximation \cite{Dai}. The electronic Hamiltonian describing the atom-ion system is discussed in Appendix \ref{App2}. The interactions of the Rydberg molecular ion is invariant with respect to exchange in nuclear positions and it is always possible and often convenient to express the molecular states in the (un)gerade basis, $|e^{\alpha,(\pm)}\rangle = 1/\sqrt{2}(|ie^{\alpha}\rangle \pm |e^{\alpha}i\rangle)$ and their corresponding energies $E^{\alpha,(\pm)}(R)$. $|ie^{\alpha}\rangle$ or $|e^{\alpha}i\rangle$ are defined depending on whether the Rydberg atom is to the right or left of the ion. Owing to the non-orthogonality between the Rydberg wave functions ($\psi_{nl}$) defined at either nuclei, there is a small but non-zero overlap function. However for this work, the focus is on inter-nuclear distances where these overlap integrals are small thereby obtaining a simplified eigenvalue problem for the electronic Hamiltonian,
\begin{equation}
\hat{H}_{\rm el} |e^{\alpha,\pm}\rangle = E^{\pm}_{\alpha} (R) |e^{\alpha,\pm}\rangle\ .
\end{equation}
The index $\alpha=1,2,\hdots$ represents the different excited states of the Rydberg molecular ion. Upon diagonalization, we have the Rydberg molecular states and energies. Compared to calculations of low-lying states, those for highly excited molecules prove very demanding due to the need for a large basis set and the highly oscillatory character of the involved atomic Rydberg states. Fig.~\ref{fig2}(a) depicts a characteristic pair of molecular potential curves $E^{\alpha,(\pm)}(R)$ around the ${\rm Sr^+_2}(50S)$ asymptote, obtained for a basis set of $\sim\!10^3$ atomic states. The numerical calculations used basis states with principal quantum number ranging from $n = 40-60$ including $l=0 \hdots (n-1)$ states for each $n$. At such high excitations, the ion-atom interaction leads to strong state mixing already at micrometer distances which is reflected in the molecular ion wave functions shown in Fig.~\ref{fig2}(b)-(d). The charge exchange between the ion and Rydberg atom is determined by the energy splitting given as 
\begin{align}
T^{\alpha}(R) = \frac{E^{\alpha,+}(R)-E^{\alpha,-}(R)}{2} \ .
\end{align}
There is substantial tunnel splitting between the opposite symmetry states [see inset of Fig.\ref{fig2}] of up to several hundred MHz, even at distances for which the Rydberg electron remains well localized at either ionic core. The polarization of the Rydberg atom due to the ion is calculated from the slope of the molecular potential curves as a function of the inter-nuclear distance.

\section{Optical coupling to Rydberg molecular ion states} 
Using the two particle notation introduced in the previous section, the ion and the ground state atom of Sr is denoted by $|ig\rangle$ or $|gi\rangle$ depending on the position of the respective particles, where $|g\rangle = |5s^2,^{1\!}\!S_0\rangle$. The optical coupling of $|ig\rangle$, $|gi\rangle$ to  $|ie^{\alpha}\rangle$, $|e^{\alpha}i\rangle$ respectively is a two photon process via the inter-mediate triplet state, 5s5p,$^{3\!}\!P_1$ with an effective Rabi frequency $\Omega^{\alpha}(R)$. The coupling is determined by the dipole matrix element, $\mu^{\alpha}(R) = \langle 5s5p|\mu|e^\alpha(R)\rangle$. In order to relate this coupling strength to that of neutral gas experiments, all Rabi frequencies are expressed in terms of a reference Rabi frequency, $\Omega_{5s}^{50s}$, for an isolated atom which for our purposes is chosen to be $40$~MHz. Detuning of the laser with respect to a particular molecular Rydberg state is given as $\Delta^{\alpha}(R)=\omega_L - (E^{\alpha,-}(R)+E^{\alpha,+}(R))/2$ where  $\omega_L$ is the frequency of the second photon. Using the dipole approximation for the laser field and the rotating wave approximation, the resulting Hamiltonian is
\begin{align}\label{Hel_new}
\hat{H}^{\rm tp}_{\rm opt}(R) =& \sum_{\alpha} \Bigg[ -\Delta^{\alpha}(R) \left(|ie^{\alpha}\rangle \langle ie^{\alpha}|+ |e^{\alpha}i\rangle \langle e^{\alpha}i| \right)  \\ \nonumber
&  + \frac{\Omega^{\alpha}(R)}{2} \left(|ig\rangle \langle ie^{\alpha}|  + |gi\rangle \langle e^{\alpha}i|  + \text{h.c.} \right) \\ \nonumber
&+ T^{\alpha}(R) \left(|ie^{\alpha}\rangle \langle e^{\alpha}i| + \text{h.c.}\right) \Bigg] \ .
\end{align}
The electronic ground states energies are set to zero. The Hamiltonian $\hat{H}^{\rm tp}_{\rm opt}$ is diagonalized to obtain exact solutions for the laser-dressed molecular states, $|d_{\beta}\rangle$ along with the energies $\omega^{\beta}(R)$ ($\hbar$ is set to 1). $|d_{\beta}\rangle$ are expressed in terms of a superposition of the electronic ground states $|ig\rangle$, $|gi\rangle$ as well as the molecular excited states $|ie^{\alpha}\rangle$, $|e^{\alpha}i\rangle$. The index $\beta=1,2,\hdots$ represents the different dressed states. Of major interest is the pair of molecular states that has the largest contribution of electronic ground states. These states correspond to states with large lifetimes and are denoted by $|\tilde{g}_{1,2}\rangle$ with energies $\omega^{\tilde{g}}_{1,2}(R)$. Expressing the electronic dynamics as effective hopping between the relevant dressed states, we have
\begin{align}\label{eo-effec}
\hat{H}^{\rm tp}_{\rm effec}(R) =& U(R) \left(|i\tilde{g}\rangle \langle i\tilde{g}|  + |\tilde{g}i\rangle\langle \tilde{g}i| \right) \nonumber \\
&+ J(R) \left(|i\tilde{g}\rangle\langle \tilde{g}i| + \text{h.c.} \right) \ .
\end{align}
where $U(R)$  is the light shift associated with ion-dressed atom pair and $J(R)$ is the effective hopping (see Fig. \ref{fig1}). The definitions of $U(R)$ and $J(R)$ contain the details of the admixture of Rydberg state to the ground state atom as determined by the laser parameters (refer to Appendix \ref{App3}).

\section{Charge dynamics with classical and quantum motion of Rydberg-dressed molecular ion} 

\begin{figure}
 	\includegraphics[width=0.99\columnwidth]{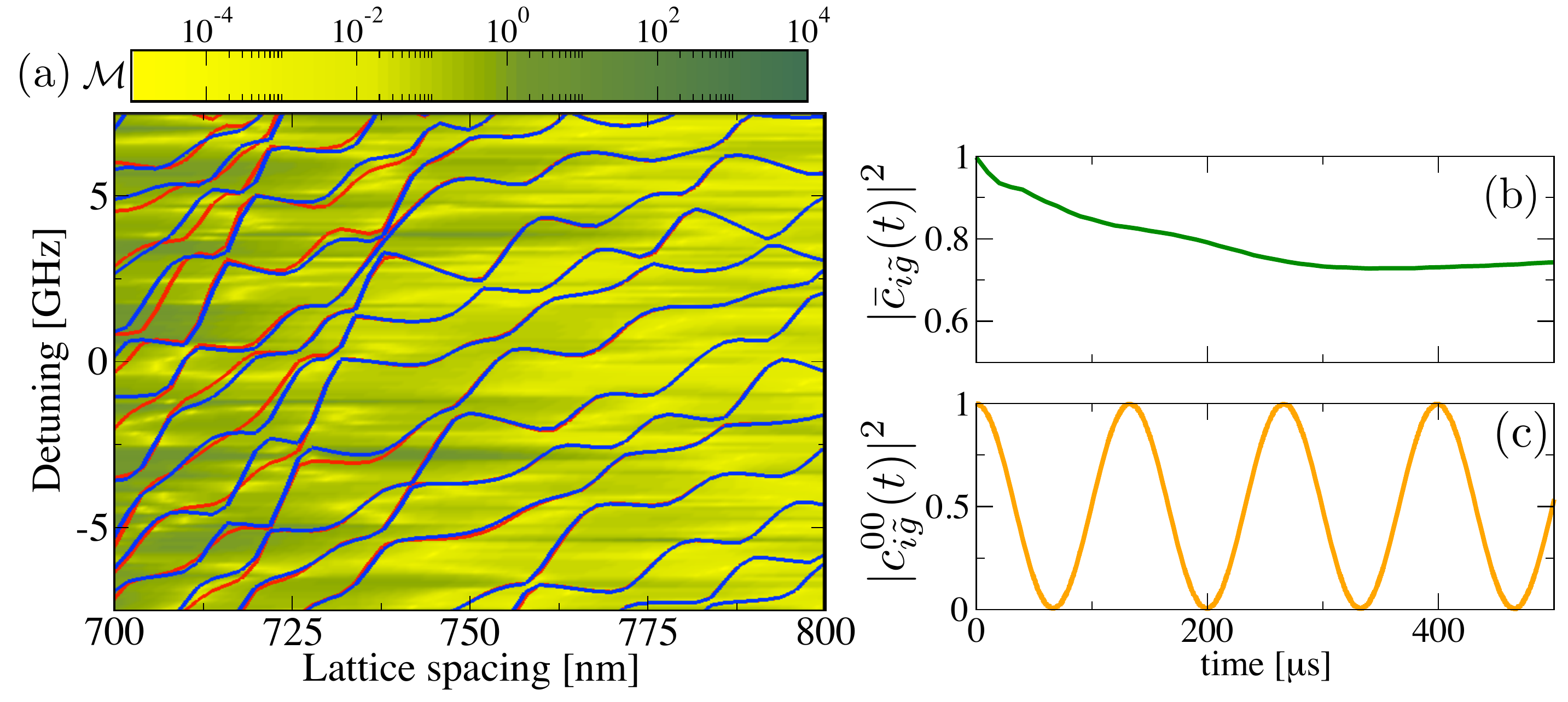}
 	\caption{(a) Colour density plot is shown for $\mathcal{M}$ (see Eq.~(\ref{Mdef})) distinguishing coherent dynamics ($\mathcal{M} \ll 1$) from incoherent dynamics ( $\mathcal{M} \ge 1$) for a range of detunings and lattice spacings. Relevant (un)gerade pair of potential curves are shown in the background as (red) blue lines for reference. (b) Plot depicts classical motional dynamics for an unconfined ion-atom pair by plotting $|\bar{c}_{i\tilde{g}}(t)|^2$ which is calculated by averaging $c_{i\tilde{g}}(R,t)$ over the inter-nuclear distance. (c) Plot shows the probability for an ion to be at a given site for an ion-atom pair trapped in a double well with trap frequency, $\omega_{\rm tr} = (2\pi) 80$ kHz and lattice spacing as 796 nm. The chosen detuning is $-0.7$ GHz which corresponds to $\mathcal{M} = 0.044$. \label{fig3}}
 \end{figure}

Here classical and quantum dynamics refer to the motional states of untrapped and trapped ion-atom pair respectively. For an unconfined pair of particles, one obtains different hopping rates corresponding to different inter-nuclear distances which leads to dephasing in the overall charge dynamics. The instantaneous state for classical dynamics is given by $|\psi(R,t)\rangle  = c_{i\tilde{g}}(R,t)|i\tilde{g}\rangle +  c_{\tilde{g}i}(R,t)|\tilde{g}i\rangle$ and the corresponding equations of motion using Eq.~(\ref{eo-effec}) are
\begin{eqnarray}\label{cig}
i\partial_t c_{i\tilde{g}}(R,t) &=& U(R) c_{i\tilde{g}}(R,t) + J(R) c_{\tilde{g}i}(R,t) \ , \\
i\partial_t c_{\tilde{g}i}(R,t) &=& U(R) c_{\tilde{g}i}(R,t) + J(R) c_{i\tilde{g}}(R,t) \ . 
\end{eqnarray}
For a fixed inter-nuclear distance $R$, one obtains the probability to be in state $|i\tilde{g}\rangle$ or $|\tilde{g}i\rangle$ to be $\cos^2[J(R)~t]$. On averaging over the inter-nuclear distance, the probability to obtain a particular two particle state, for example $|i\tilde{g}\rangle$ is given by $|\bar{c}_{i\tilde{g}}(t)|^2$  and is shown to slowly decay as seen in Fig.\ref{fig3}(b). 
In order to control the uncertainty in the position of either particle (ion/atom), we propose to have an identical confinement for the ion and atom (see Fig. 1(d)). This is achievable for alkaline-earth atoms as suggested in \cite{Mukherjee}, since the dynamic polarizability of a singly ionized alkaline-earth atom is comparable to that of a singly excited Rydberg alkaline-earth atom. In a double well, the electronic and motional degrees of freedom are entangled in the overall state given as $|\psi\rangle  = \sum_{n,n'} (c^{n,n'}_{i\tilde{g}}(t)|i\tilde{g}\rangle|n n'\rangle +  c^{n,n'}_{\tilde{g}i}(t)|\tilde{g}i\rangle|n n'\rangle)$. Here $|n n'\rangle$ is the eigenstate of the Hamiltonian corresponding to the center of mass dynamics for two particles in a double well harmonic trap \cite{Grimm} (refer to Appendix \ref{App5}). For a typical trapping frequency in the range of hundred kHz, the nuclear dynamics within the trap is much slower than the electronic dynamics and is solved under the Born-Oppenheimer approximation. It is assumed that the system is prepared in the lowest motional state denoted by $|00\rangle$. The probability to excite the first motional state can be calculated from the off-diagonal couplings $\langle 0 0 |U(R)|0 1\rangle$ and $ \langle 0 0 |J(R)|0 1\rangle$. If the off-diagonal couplings are smaller than the trapping frequency $\omega_{\rm tr}$ and the corresponding light shifts, then we have coherent dynamics. To quantify the degree to which we couple the lowest motional states to their next higher motional state, the following parameter is introduced,
\begin{equation}\label{Mdef}
\mathcal{M} = \frac{|U^{01}_{00} + J^{01}_{00}|}{|U^{01}_{01}-U^{00}_{00}+\omega_{\rm tr}|} \ , 
\end{equation}
where $A^{nm}_{n'm'}=\langle n m|A(R)|n' m'\rangle$ and $A\in \{U,J\}$. If the coupling of $|0 0 \rangle$ to $|0 1 \rangle$ is small enough then the coupling to higher motional states such as $|0 2 \rangle$ (or $|2 0 \rangle$)  are suppressed as well since they are higher order processes. Thus, the lower the value of $\mathcal{M}$, the more coherent is the charge dynamics. As expected, for sufficiently large trapping frequencies, it is possible to suppress the population of higher motional states. However there are experimental limitations to how large the optical trap frequency can be and the typical values of $(U^{01}_{01}-U^{00}_{00})$ are comparable to $\omega_{\rm tr}$. Thus to have lower values of $\mathcal{M}$, we need $|U^{01}_{00}|, |J^{01}_{00}| \ll |U^{01}_{01}-U^{00}_{00}|$, which is easily satisfied for large enough lattice spacings due to lower values of tunneling. In Fig.~\ref{fig3}(a), $\mathcal{M}$ is represented in a two dimensional colour plot for different values of detuning and lattice spacing for a fixed trapping frequency of  $\omega_{\rm tr} = (2\pi) 80$ kHz. It is always possible to find suitable lattice spacing and detuning to obtain coherent dynamics for our chosen trap frequency as done in Fig.~\ref{fig3}(c). Whenever we work in the regime where the higher motional states are not populated, we can replace $U(R)$ and $J(R)$ by $\bar{U} =U^{00}_{00}$ and $\bar{J}=J^{00}_{00}$ respectively.

\section{Generalization to many-body charge transport} 
\begin{figure}
	\includegraphics[width=0.99\columnwidth]{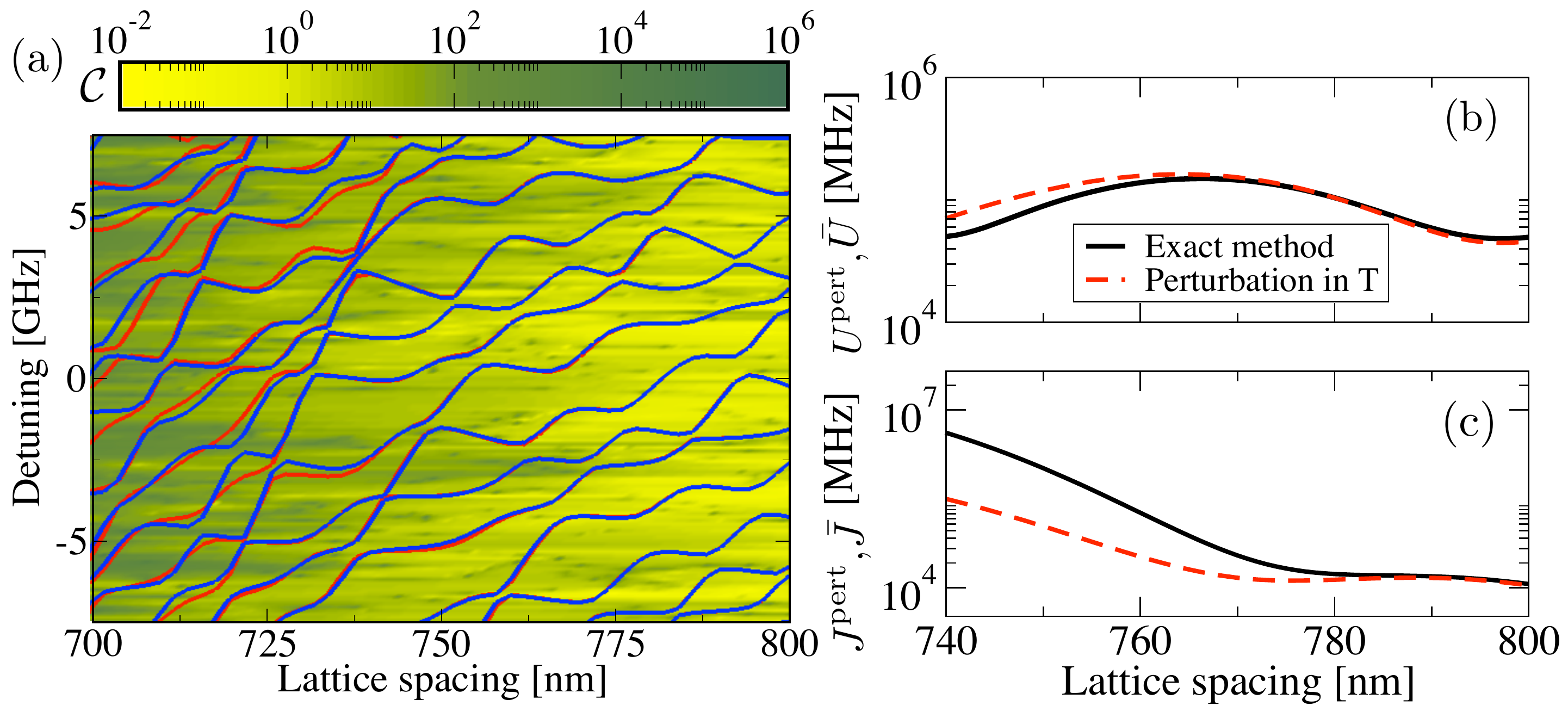}
	\caption{(a) Density plot shows the relative difference in the dynamical parameters defined in $\mathcal{C}$ (see Eq.~(\ref{Cdef})) calculated using exact as well as perturbative methods (refer to text) for different laser parameters. The light regions in the density plot correspond to regimes where the perturbation theory is valid. Relevant (un)gerade pair of potential curves are shown as (red) blue lines for reference. (b)-(c) Averaged values of $U$ and $J$ from different methods are shown for a particular detuning, $\Delta= -0.7$ GHz.\label{fig4}}
\end{figure} 
Here the two particle picture will be generalized to many particles involving a deep optical lattice filled with a single atom per site out of which one is ionized. In the many-particle system, we assume that the excitation laser parameters and lattice parameter are selectively chosen such that the atoms that are nearest neighbours to the ion have the highest probability of Rydberg excitation. We also assume that the nearest neighbouring atom on either side of the ion remain in the blockade regime despite the polarisation shifts \cite{Engel}. The shift in the energy levels is strongest for atoms that are nearest neighbours to the ion than compared to those further away from it. To this effect, we have the following reduced basis for the many-particle picture: all atoms in the ground state with an ion at site $k$ ($|I_{k}\rangle = |g_1...g_{k-1} \ i_k \ g_{k+1}...g_N\rangle $), a Rydberg atom to the right of the ion ($|R^{\alpha}_{k}\rangle =  |g_1...g_{k-1} \ i_k \ e^{\alpha}_{k+1}...g_N\rangle$) and a Rydberg atom to the left of the ion ($|L^{\alpha}_{k}\rangle = |g_1...e^{\alpha}_{k-1} \ i_k \ g_{k+1}...g_N\rangle$). Any accidental Rydberg excitation for atoms further away from the ion do not contribute to the overall ion dynamics as the tunneling rate is negligible and it couples back to $|I_k\rangle$. The Hamiltonian for the optical coupling of the many-body system is given as 
\begin{align}\label{mpopt}
\hat{H}^{\rm mp}_{\rm opt} &= \sum_{\alpha}\sum^{N}_{k=1} \left[-\Delta^{\alpha}_{k} |R^{\alpha}_{k}\rangle\langle R^{\alpha}_{k}| + \frac{\Omega^{\alpha}_{k}}{2} \left(|R^{\alpha}_{k}\rangle\langle I_{k}| + \text{h.c}  \right) \right.  \nonumber\\
&\left. -\Delta^{\alpha}_{k} |L^{\alpha}_{k}\rangle\langle L^{\alpha}_{k}| + \frac{\Omega^{\alpha}_{k}}{2} \left(|L^{\alpha}_{k}\rangle\langle I_{k}| + \text{h.c}  \right)\right] \nonumber\\
&+\sum_{\alpha} \sum^{N-1}_{k=1} \frac{T^{\alpha}_k}{2} \left[|R^{\alpha}_{k}\rangle\langle L^{\alpha}_{k+1}| + \text{h.c}  \right]  \ . 
\end{align}
On comparing with Eq.~(\ref{Hel_new}), we find that the dependence of the optical parameters and the tunneling on distance $R$ has been replaced by subscript $k$, which denotes the site number of the ion placed within the atomic lattice. Similar to the two particle picture, we can diagonalize $\hat{H}^{\rm mp}_{\rm opt}$ to obtain dressed states and focus on the many-body Rydberg-dressed ground states denoted as  $|\tilde{I}^{k}\rangle = |\tilde{g}_1...\tilde{g}_{k-1} \ i_k \ \tilde{g}_{k+1}...\tilde{g}_N\rangle$. Unlike in the two particle picture, in the many-particle setup, the electron can in principle tunnel multiple times across the lattice before it couples back to the ground state atom. Although we assume nearest neighbour tunneling and work in the reduced basis, we find that the effective ion dynamics is delocalised in the Rydberg-dressed picture. This implies that the effective equations of motion couple $|\tilde{I}_{k}\rangle$ to $|\tilde{I}_{k\pm2}\rangle$ as well and so on. Thus the effective ion dynamics in the many-body dressed atoms cannot be described simply by its nearest neighbour exchange term unless we include additional constraints. However, using time independent perturbation theory where $T_k \ll \Omega_k$ for all $k$, it is possible to derive the effective nearest neighbour hopping term. The effective Hamiltonian obtained in this limit describes charge dynamics for an ion in Rydberg dressed atomic lattice,
\begin{align}\label{Hmp}
\hat{H}^{\rm mp} _{\rm effec} &= \sum_k U_{k} \left(|\tilde{I}_{k}\rangle \langle \tilde{I}_{k}| \right) \nonumber \\
&+ J_{k,k+1} \left(|\tilde{I}_{k+1}\rangle\langle |\tilde{I}_{k}| + \text{h.c.} \right) \nonumber \\
&+ J_{k,k-1} \left(|\tilde{I}_{k-1}\rangle\langle |\tilde{I}_{k}| + \text{h.c.} \right) \ ,
\end{align}
where $U_{k}  = \langle \tilde{I}_{k}|\hat{H}^{\rm mp} _{\rm effec} |\tilde{I}_{k}\rangle$ and $J_{k,k+1}  = \langle \tilde{I}_{k}|\hat{H}^{\rm mp} _{\rm effec} |\tilde{I}_{k+1}\rangle$.To identify regimes in the parameter space where the perturbation theory is valid we resort back to the  two particle picture. We derive dynamical parameters ($U^{\rm pert},J^{\rm pert}$) by solving $\hat{H}^{\rm tp}_{\rm opt}$ perturbatively in the limit $\Omega \gg T$. Averaging over the motional states, $U^{\rm pert},J^{\rm pert}$ are compared to $\bar{U},\bar{J}$ which were obtained by solving Eq.~(\ref{eo-effec}) without any approximation (referred to as exact method in Fig.~\ref{fig4}(b)-(c)). This is numerically quantified by the following parameter,
\begin{equation}\label{Cdef}
\mathcal{C} = |\frac{\bar{U}-U^{\rm pert}}{\bar{U}}| + |\frac{\bar{J}-J^{\rm pert}}{\bar{J}}| .
\end{equation}
Fig.~\ref{fig4}(a) shows the different values of $\mathcal{C}$ for different laser parameters and lattice spacing. As expected, for larger lattice spacings, the overall $T$ is smaller which easily satisfies our condition for perturbation theory and corresponds to lower values of $\mathcal{C}$. This is further confirmed in Fig.~\ref{fig4}(b)-(c) where we compare the dynamical parameters from two different methods. We include the motional states in next section, where we work in the coherent regime of the many-body setup.

\section{Coherent many-body charge transport with nearest-neighbour hopping} 
 \begin{figure}
 	\includegraphics[width=0.99\columnwidth]{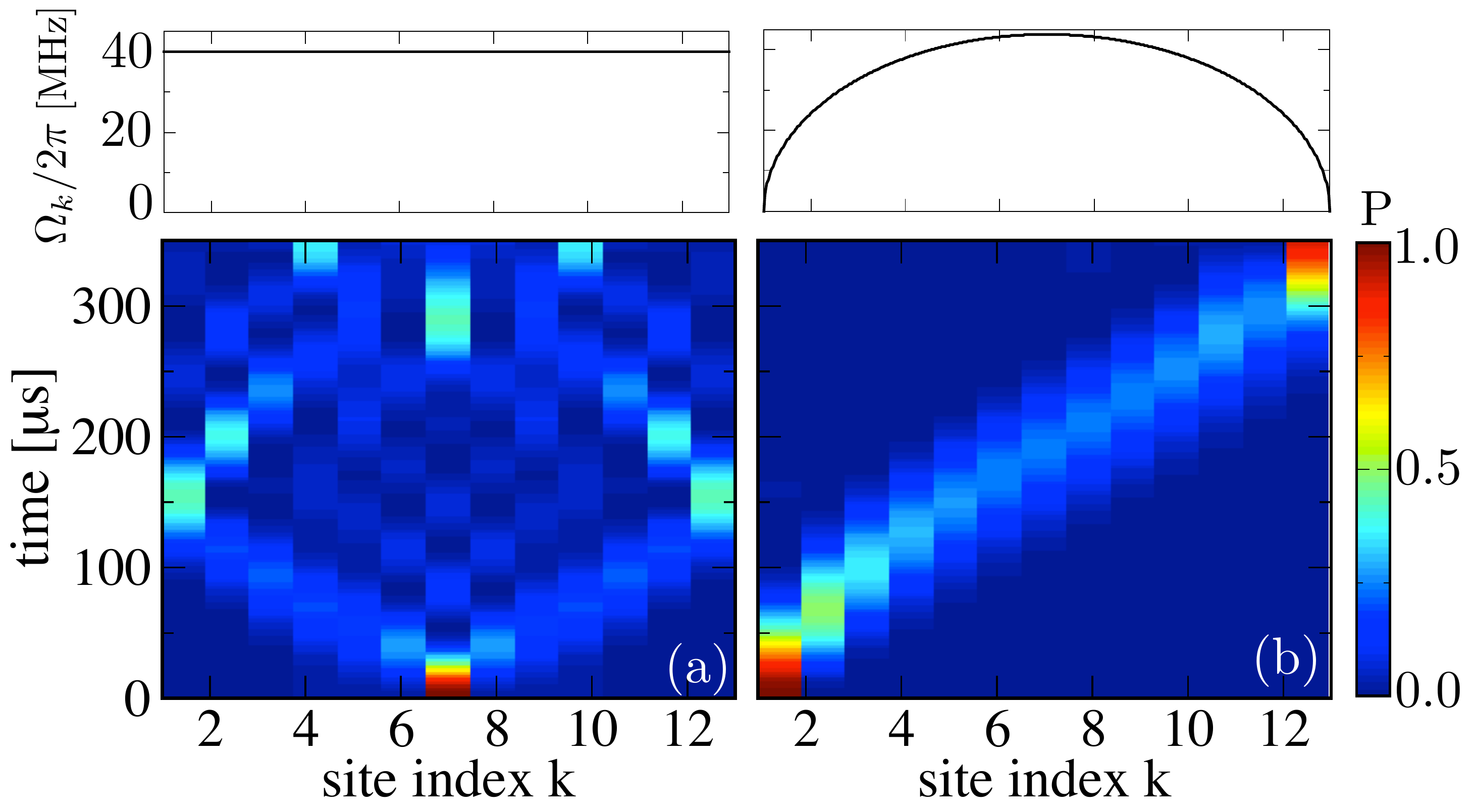}
 	\caption{Density plot showing ion probability during its dynamics in Rydberg dressed lattice with a spacing of 796 nm and laser detuning -0.7 GHz. Top panels are showing the corresponding Rabi profile of the excitation laser: (a) Constant Rabi frequency, $\Omega_c = 40$ MHz, where $J_0=\bar{J}_{k,k+1} = 27$ kHz and $U_0=\bar{U}_k = 15.9$ kHz for all values of $k$. (b) Varying Rabi profile along the x-direction, $\Omega_k(x) = \Omega_c \sqrt{(N - kx)(k-1)x}$ such that $\bar{J}_{k,k+1} =(2 J_0/N\sqrt{(N-k)k}$ and $\bar{U}_{k,k+1} =(2 U_0/N\sqrt{(N-k)k}$. \label{fig5}}
 \end{figure}
The theory for charge transport over many-sites can be understood using the simple model of pair-wise charge exchange at two sites involving the ion and its neighbouring atom. Having identified the optimum optical parameters ($0.7-1$ GHz with respect to $50s$ Rydberg state) and lattice spacing (750-850 nm) in prior sections to have nearest neighbour coherent hopping, we use them in our numerical simulation for charge dynamics involving a single ion and $N-1$ atoms in a one dimensional optical lattice. The typical lifetime of the Rydberg states is estimated to be in the order of hundred $\mu$s \cite{Gallagher1994} which is enhanced to 250-350 ms by averaging over the Rydberg dressed states whose major contribution is from electronic ground state. Thus the main time constraint on the overall dynamics is the decay of the inter-mediate state, 5s5p,$^{3\!}\!P_1$, which is 21 $\mu$s. By increasing the detuning with respect to the inter-mediate state, this is increased to 8.4 ms. This requires the effective Rabi frequency for the two photon excitation scheme to be in the order of tens of MHz, which remains experimentally achievable \cite{Millen2010,mcquillen13,lochead13}. We ignore the accidental resonances to Rydberg states or doubly excited states since the probability for it to occur is small ($\sim 10^{-2}$) particularly when averaged over the motional states. Solving for $\Psi = \sum^N_{k=1} \tilde{C}^{I}_{k} |\tilde{I}_{k}\rangle$ using the many-body Hamiltonian Eq.~(\ref{Hmp}) gives
\begin{equation}\label{manybodyeq}
i\partial_t \tilde{C}^{I}_{k} = \bar{U}_{k}\tilde{C}^{I}_{k} + \bar{J}_{k,k-1} \tilde{C}^{I}_{k-1} + \bar{J}_{k,k+1} \tilde{C}^{I}_{k+1} \ , 
\end{equation}
where $\bar{U}_{k} =\langle 00|U_k|00\rangle$ and  $\bar{J}_{k,k+1} =\langle 00|J_{k,k+1}|00\rangle$. In general, $\bar{J}_{k,k+1}$ does not have to be equal to $\bar{J}_{k+1,k}$. Fig.~\ref{fig5} depicts the results of the numerical simulation for  13 sites with different laser profiles. Focusing on the excitation laser with constant Rabi profile [see Fig.\ref{fig5}(a)], we have an ion initially located at site 7 which then propagates in both directions symmetrically as it has equal probability to hop in either direction at every instant. A scenario involving spatially varying Rabi profile is depicted in Fig.\ref{fig5}(b). In this case $\bar{J}_{k,k+1} \neq \bar{J}_{k+1,k}$ and the Rabi profile has been chosen in such a manner that it mimics motion of a particle in harmonic well \cite{Christandl} with its minima at the center (site 7 in this case). Hence an ion situated at site 1 is akin to starting at one end of the well which then propagates through site 7 with maximum kinetic energy till it reaches the other end. The role of the light shift $\bar{U}_k$ is simply an additional energy shift experienced by the atoms/ion in the lattice. This can be compensated by choosing an appropriate profile for the trapping laser.

\section{Conclusion} 

In this work, we propose and model the effective charge dynamics of an ion within trapped Sr atoms in an optical lattice. We conclude that optically trapped alkaline-earth atoms-ion systems can naturally serve as a platform for the study of charge transfer in a controlled many-body environment. Enhanced coherent charge dynamics requires the dressing of ground state atoms to their Rydberg states \cite{Gil,Mukherjee2} and the provision of identical confinement for both the ion and the Rydberg-dressed atom \cite{Mukherjee}, both of which are potentially attainable with ongoing experiments with alkaline-earth atoms \cite{Snigirev,Bounds,Norcia, Cooper,Couturier,Hu}. Recent experiments on optical trapping of ions \cite{Schmidt2}, ion-Rydberg atoms \cite{Ewald,Engel} and ion-dressed Rydberg atoms \cite{Secker} are all promising endeavors in realising different aspects of this work.

R.M would like to acknowledge I. Lesanovsky and T. Pohl for their invaluable input and discussion. R.M would also like to acknowledge S. W\"{u}ster for his discussion and the Max-Planck society for funding under the MPG- IISER partner group program.

\appendix


\section{Full Hamiltonian}\label{App1}

In the simple picture of two particles where a Sr atom is next to an ion, the full Hamiltonian consists of three parts,
\begin{equation}\label{fullH}
	\hat{H} = (\hat{H}_{\rm el} + \hat{H}^{\rm tp}_{\rm opt} + \hat{H}_{\rm CoM} ) ,
\end{equation}
where $\hat{H}_{\rm el}$ is the electronic part, $\hat{H}^{\rm tp}_{\rm opt}$ represents the excitation of the atom to its Rydberg state and $\hat{H}_{\rm CoM}$ corresponds to the motion of the trapped particles in the lattice. We discuss each Hamiltonian in some detail in the following sections.

\section{Electronic Hamiltonian}\label{App2}

Similar to alkali atoms, we assume an effective model potential for the singly ionized alkaline-earth atom, reducing the many electron problem to an effective two electron atom problem. The model potential in atomic units is given as
\begin{align}
	V^{Sr^{+}}_{eff}(r) &= -\frac{1}{r}[2+(Z-2) e^{-a_{1}(l) r} + a_{2}(l) r e^{-a_{3}(l) r}] \nonumber \\
	&-\frac{\alpha_c}{2r^4}[1 - e^{-(r/r_l)^6}]  ,
\end{align} 
where parameters $(a_1(l) ,a_2(l) ,a_3(l) ,\alpha_c,r_l)$ of the model potential are determined from fits to experimental data for low and inter-mediate levels of Sr$^{+}$ energies \cite{Guet}. $l$ is the orbital angular momentum. Using the singly ionized Rydberg wavefunctions as a basis, the atomic Sr Rydberg wavefunctions $\psi_{n,l}(\mathbf{r})$ were calculated using mean field theory similar to Hartree-Fock theory. The mixing between Rydberg series are ignored and is justified in certain cases from experimental observations which show no high lying perturbers in the Rydberg series. 

\begin{figure}[t!]
	\includegraphics[width=0.6\columnwidth]{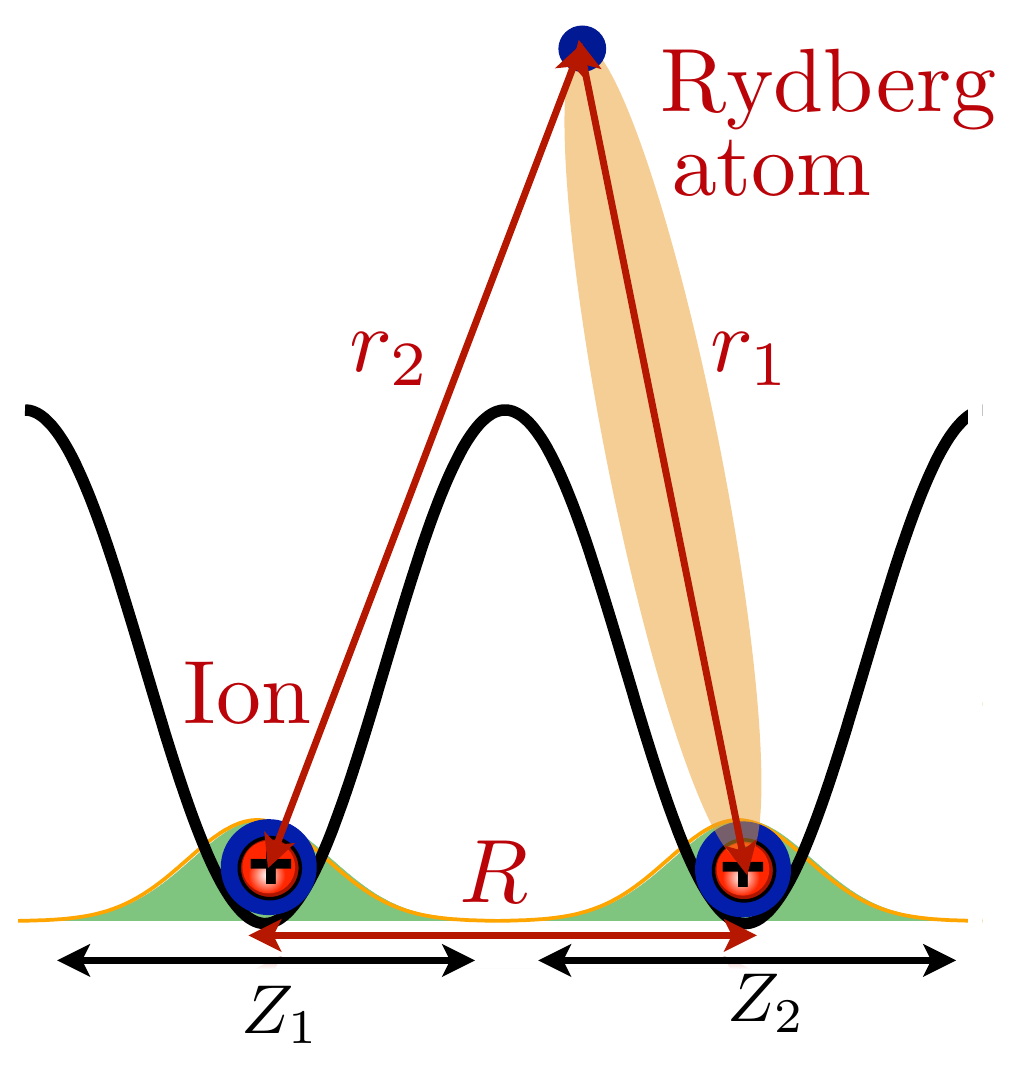}
	\caption{The figure depicts a Sr Rydberg molecular ion trapped in a double well. Here $r_{i=1,2}$ is the relative position of the Rydberg electron with respect to either nucleus and $R$ is the inter-nuclear distance between the nuclei. $Z_i$ is the relative displacement of the corresponding nucleus with respect to the center of lattice site $i$.} \label{Appfig}
\end{figure}

For singly excited Rydberg states of strontium with very large principal quantum numbers ($n>20$), there is a large asymmetry in the orbit size of the Rydberg electron and the ground state electron. Due to this asymmetry, we can treat the exchange interaction between the two valence electrons perturbatively. Since the effect of the inner valence electron (in its ground state) is negligible on the Rydberg electron, we write the electronic Hamiltonian in terms of the Rydberg electron, given as 
\begin{equation}
	\hat{H}_{\rm el} = -\frac{\bigtriangledown^2}{2} + V^{Sr^{+}}_{eff}(\mathbf{r_1}) + V^{Sr^{+}}_{eff}(\mathbf{r_2}) + \frac{1}{\mathbf{R}} ,
\end{equation}
where $r_{i=1,2}$ is the relative position of the Rydberg electron with respect to either nucleus (see Fig.\ref{Appfig}) and $R$ is the inter-nuclear distance between the nuclei. We note that the basis states are not orthonormal given the small non-zero overlap function i.e. $\langle \psi_{n,l}(\mathbf{r_{(1,2)}})|\psi_{n,l}(\mathbf{r_{(2,1)}})\rangle \neq 0$. However for inter-nuclear distances considered in this work, the overlap function can be considered negligible.  
We solve the  (approximate) eigenvalue problem by diagonalising the following matrix Hamiltonian for $\hat{H}_{\rm el}$,  
\begin{equation}
	\begin{pmatrix}
		[P(r_{1})]  & [T(r_1,r_2)] \\
		[T(r_2,r_1)]  & [P(r_{2})] 
	\end{pmatrix} 
\end{equation}
where $[P(r_{i})]$ and $[T(r_i,r_j)]$ are block matrices given as
\begin{equation}
	[P(r_{i})] =  
	\begin{pmatrix}
		P^{nl}_{nl}(r_{i})  & P^{n'l'}_{nl}(r_{i})  & \cdots \\
		P^{nl}_{n'l'}(r_{i}) & \ddots & \cdots \\
		\vdots & \cdots & \ddots \\
	\end{pmatrix} 
\end{equation}

\begin{equation}
	[T(r_i,r_j)] =  
	\begin{pmatrix}
		T^{nl}_{nl}(r_i,r_j) & T^{n'l'}_{nl}(r_i,r_j) & \cdots\\
		T^{nl}_{n'l'}(r_i,r_j) & \ddots & \vdots \\
		\vdots & \vdots & \vdots \\
	\end{pmatrix} 
\end{equation}
which are defined using the two center integrals $P^{nl}_{n'l'}(r_{i=1,2})=\langle\psi_{n,l}(\mathbf{r_{i=1,2}})|\hat{H}_{\rm el}|\psi_{n',l'}(\mathbf{r_{i=1,2}})\rangle$ and $T^{nl}_{n'l'}(r_1,r_2)=\langle \psi_{n,l}(\mathbf{r_{(1,2)}})|\hat{H}_{\rm el}|\psi_{n',l'}(\mathbf{r_{(2,1)}})\rangle$.  Eq.~(4) given above can be a very large matrix to diagonalise for a large basis set and needs to be computed for inter-nuclear distances that are finely resolved to obtain smooth potential curves. This is numerically cumbersome. However, using the `symmetry' between $\mathbf{r_1}$ and $\mathbf{r_2}$ in the electronic Hamiltonian, we express the Hamiltonian in the \textit{gerade-ungerade} basis representation. The advantage of this representation is that the set of eigenvalue equations de-couple between the states belonging to the two symmetry groups. Upon diagonalization, one obtains the following eigenvalue equation for the Rydberg states,
\begin{equation}
	\hat{H}_{\rm el} |e^{\alpha,\pm}\rangle = E^{\pm}_{\alpha} (R) |e^{\alpha,\pm}\rangle ,
\end{equation}
where $\alpha$ is the index for the Rydberg excited molecular ion state which in terms of the basis functions is expressed as
\begin{align}
	|e^{\alpha,\pm}\rangle &= \frac{1}{\sqrt{2}}(|ie^{\alpha}\rangle \pm |e^{\alpha}i\rangle) \nonumber \\
	&= \sum_{n,l} c^{\alpha,\pm}_{n,l} \left(|\psi_{n,l}(\mathbf{r_1})\rangle  \pm |\psi_{n,l}(\mathbf{r_2})\rangle  \right) ,
\end{align}
along with the normalization condition,$\sum_{n,l} |c^{\alpha,\pm}_{n,l}|^2 = 1$.

\section{Hamiltonian for optical coupling to Rydberg molecular ion states}\label{App3}

The laser-dressed molecular ion states are obtained by diagonalising the following matrix Hamiltonian of $\hat{H}^{\rm tp}_{\rm opt}$,
\begin{equation}\label{origpertmat}
	\begin{pmatrix}
		0 & \frac{\Omega^{\alpha}(R)}{2} & \hdots & 0 & 0 &\hdots\\[5pt]
		\frac{\Omega^{\alpha}(R)}{2} & -\Delta^{\alpha}(R) & \vdots & 0 & T^{\alpha}(R) & \vdots \\[5pt]
		\vdots & \hdots & \ddots & \vdots & \hdots & \ddots \\[5pt]
		0 & 0 & \hdots & 0 & \frac{\Omega^{\alpha}(R)}{2} & \hdots \\[5pt]
		0 & T^{\alpha}(R) & \vdots & \frac{\Omega^{\alpha}(R)}{2} & -\Delta^{\alpha}(R)& \vdots \\[5pt]
		\vdots & \hdots & \ddots & \vdots & \hdots & \ddots \\[5pt]
	\end{pmatrix}.
\end{equation}
The size of the matrix Hamiltonian is $(2+\alpha)\times (2+\alpha)$ where $\alpha$ is the number of Rydberg molecular ion states. The eigenvalues obatined $\omega^{\beta}(R)$ were ordered in ascending magnitude of energies along with their corresponding eigenstates,
\begin{align}
	|d_{\beta}(R) \rangle &= c^{g_1}_{\beta}(t) |g_1\rangle + c^{g_2}_{\beta}(t)  |g_2\rangle + \nonumber \\ 
	&\sum_{\alpha=1} \left(c^{e_1,\alpha}_{\beta}(t)  |e^{\alpha}_1\rangle + c^{e_2,\alpha}_{\beta}(t)  |e^{\alpha}_2\rangle  \right) ,
\end{align}
where $\beta$ represents different dressed molecular ion states. The states with lowest energy in magnitude are selected states and denoted as $|\tilde{g}_{1,2}\rangle$ with energies $\omega^{\tilde{g}}_{1,2}(R)$. These states would correspond to states with longest lifetimes depending on the fraction of the Rydberg population in them which in turn depends on the Rabi frequency and the detuning of the excitation laser. 


\section{Electronic dynamics in the Rydberg dressed molecular ion}\label{App4}

We solve the time dependent Schr\"{o}dinger equation for $|\Psi\rangle$,
\begin{equation}
	i\frac{\rm d |\Psi \rangle}{\rm dt} = \left(\hat{H}_{\rm el}+\hat{H}^{\rm tp}_{\rm opt}\right) |\Psi\rangle ,
\end{equation}
where  $|\Psi\rangle = c^{\tilde{g}}_{1}(R,t) |\tilde{g}_{1}\rangle + c^{\tilde{g}}_{2}(R,t) |\tilde{g}_{2}\rangle$ is expressed in terms of the Rydberg dressed ground states $|\tilde{g}_{1,2}(R)\rangle$ obtained in the previous section. The dynamical equations for the electron between the Rydberg dressed ground states are
\begin{align}\label{numgerunger}
	i \partial_t c^{\tilde{g}}_{1}(R,t)  &= \omega^{\tilde{g}}_{1}(R) c^{\tilde{g}}_{1}(R,t) \ , \\
	i \partial_t c^{\tilde{g}}_{2}(R,t)  &= \omega^{\tilde{g}}_{2}(R) c^{\tilde{g}}_{2}(R,t) \ .
\end{align}
The dressed states expressed in the left/right basis are defined as
\begin{align}
	|i\tilde{g}\rangle &= \frac{1}{\sqrt{2}} \left(\tilde{g}_{1} + \tilde{g}_{2} \right) \ , \\
	|\tilde{g}i\rangle &= \frac{1}{\sqrt{2}} \left(\tilde{g}_{1} - \tilde{g}_{2} \right) \ .
\end{align}
Using $|\Psi\rangle = c_{i\tilde{g}}(R,t) |i\tilde{g}\rangle + c_{\tilde{g}i}(R,t) |\tilde{g}i\rangle$ we have the following dynamical equations in the left/right basis,
\begin{align}
	i \partial_t c_{i\tilde{g}}(R,t) &= U(R) c_{i\tilde{g}}(R,t) + J(R) c_{\tilde{g}i}(R,t) \ ,\\
	i \partial_t c_{\tilde{g}i}(R,t) &= U(R) c_{\tilde{g}i}(R,t) + J(R) c_{i\tilde{g}}(R,t) \ ,
\end{align}
where the dynamical parameters, on-site energy $U(R)$ and the hopping rate $J(R)$ are defined as
\begin{align}
	U(R) &= \frac{\omega^{\tilde{g}}_{1}(R)+\omega^{\tilde{g}}_{2}(R)}{2} , \\
	J(R) &= \frac{\omega^{\tilde{g}}_{1}(R)-\omega^{\tilde{g}}_{2}(R)}{2} .
\end{align}
$U(R)$ and $J(R)$ are the dynamical parameters that correspond to the light shift and the effective hopping rate respectively between the Rydberg dressed ground states for an ion and an atom.  

\section{Hamiltonian for center of mass dynamics}\label{App5}

Each ion core is trapped in its own lattice site in a 1D lattice along the $z$ axis (see Fig. \ref{Appfig}). The Hamiltonian for the nuclear motion for our two site model is given as
\begin{align}\label{chargecom}
	\hat{H}_{\rm CoM} |n_1 n_2\rangle &= \sum_{i=1,2} \left[-\frac{\hbar^2\bigtriangledown_{Z_{i}}^2}{2M} + \frac{M}{2}\omega^2 Z_i^2\right] |n_1 n_2\rangle \nonumber \\
	&= \hbar(n_1 +n_2)\omega |n_1 n_2\rangle \ ,
\end{align}   
where $M =87.2$ is the mass of Sr in atomic units. and $|n_{i=1,2}\rangle$ is the motional state at the corresponding site and $|n_1 n_2\rangle$ is the two particle motional eigenstate.$Z_{(1,2)}$ is the relative motion of the corresponding trapped nuclei at each site. The full wavefunction $|\Psi\rangle$ is a product state of the electronic eigenstates and the motional states,
\begin{equation}
	|\Psi \rangle = \sum_{n_1,n_2} \left(c^{i\tilde{g}}_{n_1 n_2}(t)~|i\tilde{g}\rangle+ c^{\tilde{g}i}_{n_1 n_2}(t)~|\tilde{g}i\rangle \right) |n_1 n_2\rangle \ . \label{Chpion:Harwf} 
\end{equation}
and solving the Schr\"{o}dinger equation with the full Hamiltonian (see Eq.~(\ref{fullH})) and multiplying with $\langle n_1 n_2|$ throughout we get
\begin{align}
	i\partial_t c^{i\tilde{g}}_{n_1 n_2} =& \left(\langle n_1 n_2|U(R)|n_1 n_2\rangle + \omega_{n_1 n_2}\right) c^{i\tilde{g}}_{n_1 n_2} \nonumber\\
	&+ \left(\langle n_1 n_2|J(R)|n_1 n_2\rangle\right) c^{\tilde{g}i}_{n_1 n_2} \nonumber\\
	&+ \sum_{n_1,n_2\neq n'_1 n'_2}\left[\langle n'_1 n'_2|U(R)|n_1 n_2\rangle \right] c^{i\tilde{g}}_{n'_2 n'_2} \nonumber\\
	&+ \sum_{n_1,n_2\neq n'_1 n'_2}\left[\langle n'_1 n'_2|J(R)|n_1 n_2\rangle \right] c^{\tilde{g}i}_{n'_2 n'_2} \ , \\ 
	i\partial_t c^{\tilde{g}i}_{n_1 n_2} =& \left(\langle n_1 n_2|U(R)|n_1 n_2\rangle + \omega_{n_1 n_2}\right) c^{\tilde{g}i}_{n_1 n_2} \nonumber \\
	&+ \left(\langle n_1 n_2|J(R)|n_1 n_2\rangle\right) c^{i\tilde{g}}_{n_1 n_2} \nonumber \\
	&+ \sum_{n_1,n_2\neq n'_1 n'_2}\left[\langle n'_1 n'_2|U(R)|n_1 n_2\rangle \right] c^{\tilde{g}i}_{n'_2 n'_2} \nonumber \\
	&+ \sum_{n_1,n_2\neq n'_1 n'_2}\left[\langle n'_1 n'_2|J(R)|n_1 n_2\rangle \right] c^{i\tilde{g}}_{n'_2 n'_2} \ . 
\end{align}
On averaging over the spatial variation of the dynamical parameters, we get couplings between the motional ground states but also to other higher motional states. The criterion for coherent charge dynamics is discussed in the article and deals with finding optimal parameters that minimizes this coupling to higher motional states. Here we also numerically verified that the variation of the Rydberg dressed ground state wavefunction over the inter-nuclear distance is negligible.

\end{document}